\documentstyle[preprint,aps]{revtex}

\begin{document}
\title{Identical Relations among Transverse Parts of Variant Green Functions and the Full Vertices in Gauge Theories}
\author{Han-xin He}
\address{
China Institute of Atomic Energy, P.O.Box 275(18),Beijing 102413 P.R.China\\
and Department of Physics, Theoretical Physics Institute, University of Alberta, Canada}

\maketitle
\begin{abstract}
The identical relations among the transverse parts of variant vertex functions are derived by computing the curl of the time-ordered products of three-point Green functions involving the vector, the axial-vector and the tensor current operators, respectively. These transverse relations are coupled each other. Combining these transverse relations with the normal (longitudinal)
Ward-Takahashi identities forms a complete set of constraint relations for three-point vertex functions. As a consequence, the full vector, the full axial-vector and the full tensor vertex functions in the momentum space are exactly obtained.

\end{abstract}

\newpage
In quantum field theory symmetries lead to relations among the Green functions of the theory, which are referred to as the Ward-Takahashi(W-T) identities[1]. They play an important role in proving renormalizability and in providing a consistent description in the perturbation approach of any quantum field theory. But the normal W-T identities specify only the longitudinal part of Green functions, leaving the transverse part undetermined[2].
Therefore, to obtain the complete constraint on the vertex functions and
 then to obtain the full vertex functions we need to study the W-T type's constraint relations for the transverse parts of variant vertices, which is of great significance. In this regard, an very interesting problem relates to the Dyson-Schwinger equation(DSE) approach[3].

The Dyson-Schwinger equations embody the full structure of any field theory
and consequently provide the natural way to study the dynamics such as describing the dynamical chiral symmetry breaking, confinement and other problems of hadronic physics[4]. The structure of DSEs is such that they relate the n-point Green function to the (n+1)-point function; at its simplest, propagators are related to three-point vertices, thus leading to an infinite set of coupled equations. Therefore, one has to find some way to truncate this set of equations. If we can express the full three-point vertices in terms of the two-point functions, these equations will form a closed system for the two-point functions. How to solve exactly the transverse part of the vertex and so the full vertex function then becomes a crucial and open problem [2]. Up to now this open problem has not been solved yet. Although there have
been several attempts to construct the transverse part of the vertex by Ans\"{a}tze which satisfies some constraints[2][5], however, all of Ans\"{a}tze remain {\sl ad hoc} but without considering the constraint imposed by the symmetry of the system. The latter is the key point to understand the transverse part of the vertex as in the case of the longitudinal part of the vertex.

In this letter we first present the W-T type's identical relations among the transverse parts of variant three-point vertex functions in gauge theories, which are derived by computing the curl of the time-ordered products of three-point Green functions involving the vector, the axial-vector and the tensor current operators, respectively[6], and then derive the transverse and the full vertices
by using these transverse and the normal (longitudinal) W-T relations. This approach is motivated by the fact that the normal W-T identities which specify the longitudinal part of Green functions have been derived by computing the divergence of the time-ordered products of corresponding Green functions[7]. We find three sets of transverse W-T type's relations for the vector, the axial-vector and the tensor vertex functions, respectively,
 which are coupled each other. These relations are given in the coordinate space as well as in the momentum space. The latter form is partically useful. Combining these transverse relations with the normal (longitudinal) W-T identities for the vector, the axial-vector and the tensor vertex functions leads a complete set of W-T type's constraint relations for the fermion's three-point functions. As a consequence, the full vector, the full axial-vector and the full tensor vertex functions in the momentum space are 
then consistently and exactly deduced by solving this complete set of W-T
relations in the momentum space without any Ans\"{a}tze.

Let us first briefly describe the basic approach of computing the curl of the time-ordered products of the fermion's three-point functions involving the vector, the axial-vector and the tensor current operators, respectively. For the convenience, we introduce three bilinear covariant current operators:
$V^{\lambda \mu \nu }(x)=\frac 12\bar{\psi}(x)[\gamma ^\lambda ,\sigma ^{\mu \nu }]\psi (x)=i(g^{\lambda \mu }j^\nu (x) -g^{\lambda \nu }j^\mu  (x))$ ,
$V_5^{\lambda \mu \nu }(x)=\frac 12\bar{\psi}(x)[\gamma ^\lambda ,\sigma ^{\mu \nu }]\gamma _5\psi (x)=i(g^{\lambda \mu }j_5^\nu (x) -g^{\lambda \nu }j_5^\mu (x))$, and
$V^{\lambda \mu \nu \alpha }(x)=\frac 14\bar{\psi}(x)([\gamma ^\lambda ,\sigma ^{\mu \nu }]\gamma ^\alpha -\gamma ^\alpha [\gamma ^\lambda ,\sigma ^{\mu \nu }])\psi (x)=g^{\lambda \mu }j^{\nu \alpha}(x) -g^{\lambda \nu }j^{\mu \alpha}(x)$,
 where $j^\mu (x)=\bar{\psi}(x)\gamma ^\mu \psi (x)$ , $j_5^\mu (x)=\bar{\psi}(x)\gamma ^\mu \gamma _5\psi (x)$, and
$j^{\mu \nu }(x)=\bar{\psi}(x)\sigma ^{\mu \nu }\psi (x)$. 
 Thus the curl of the T-product of the corresponding fermion's three-point function is given by $\partial _\lambda ^xT(V^{\lambda \mu \nu }(x)\psi (x_1)\bar{\psi}(x_2))$ or 
$\partial _\lambda ^xT(V_5^{\lambda \mu \nu }(x)\psi (x_1)\bar{\psi}(x_2))$
or $\partial _\lambda ^xT(V^{\lambda \mu \nu \alpha}(x)\psi (x_1)\bar{\psi}(x_2))$, 
where $\partial _\lambda ^x$ denotes the derivative operator with respect to the argument $x$. In terms of the definition for the time-ordered product of fermion fields and the equal-time anticommutation relations for fermion fields, it is not difficult to carry out the above differential operations. The procedure is similar to that for deriving the normal W-T identities[7]. We thus find the following covariant identical relations in the operator form:
\begin{eqnarray}
& &\partial _x^\mu T(j^\nu (x)\psi (x_1)\bar{\psi}(x_2))-\partial _x^\nu T(j^\mu (x)\psi (x_1)\bar{\psi}(x_2)) \nonumber \\
&=& i\sigma ^{\mu \nu }T(\psi (x_1)\bar{\psi}(x_2))\delta ^4(x_1-x) +iT(\psi (x_1)\bar{\psi}(x_2))\sigma ^{\mu \nu }\delta ^4(x_2-x)\nonumber\\
& &+T(\bar{\psi}(x)(\sigma ^{\mu \nu }i\stackrel{\rightarrow}{\makebox[-0.8 mm][l]{/}{D}}_x-i\stackrel{\leftarrow}{\makebox[-0.8 mm][l]{/}{D}}_x\sigma ^{\mu \nu })\psi (x)\psi (x_1)\bar{\psi}(x_2))\nonumber\\
& &+{\lim _{x^{\prime }\rightarrow x}}i(\partial _\lambda ^x-\partial _\lambda ^{x^{\prime }})T(\bar{\psi}(x^{\prime })\varepsilon ^{\lambda \mu \nu \rho }\gamma _\rho \gamma _5U_P (x^{\prime },x)\psi (x)\psi (x_1)\bar{\psi}(x_2)),
\end{eqnarray}
\begin{eqnarray}
& &\partial _x^\mu T(j_5^\nu (x)\psi (x_1)\bar{\psi}(x_2))-\partial _x^\nu T(j_5^\mu (x)\psi (x_1)\bar{\psi}(x_2))\nonumber \\
&=&i\sigma ^{\mu \nu }\gamma _5T(\psi (x_1)\bar{\psi}(x_2))\delta ^4(x_1-x)-iT(\psi (x_1)\bar{\psi}(x_2))\sigma ^{\mu \nu }\gamma _5\delta ^4(x_2-x)\nonumber \\
& &-T(\bar{\psi}(x)(i\stackrel{\leftarrow}{\makebox[-0.8 mm][l]{/}{D}}_x\sigma ^{\mu \nu }\gamma _5+i\sigma ^{\mu \nu }\gamma _5\stackrel{\rightarrow}{\makebox[-0.8 mm][l]{/}{D}}_x)\psi (x)\psi (x_1)\bar{\psi}(x_2)) \nonumber \\
& &+{\lim _{x^{\prime }\rightarrow x}}i(\partial _\lambda ^x-\partial _\lambda ^{x^{\prime }})T(\bar{\psi}(x^{\prime })\varepsilon ^{\lambda \mu \nu \rho }\gamma _\rho U_P (x^{\prime },x)\psi (x)\psi (x_1)\bar{\psi}(x_2)),
\end{eqnarray}
and
\begin{eqnarray}
& &\partial _x^\mu T(j^{\nu \alpha}(x)\psi (x_1)\bar{\psi}(x_2))-\partial _x^\nu T(j^{\mu \alpha}(x)\psi (x_1)\bar{\psi}(x_2))\nonumber \\
&=&\varepsilon ^{\mu \nu \alpha \rho}\gamma _\rho \gamma_5T(\psi(x_1)
\bar{\psi}(x_2))\delta ^4(x_1-x)-T(\psi (x_1)\bar{\psi}(x_2))\varepsilon ^{\mu \nu \alpha \rho}\gamma_\rho \gamma _5\delta ^4(x_2-x)\nonumber \\
& &-T(\bar{\psi}(x)\varepsilon ^{\mu \nu \alpha \rho }
(\stackrel{\leftarrow}{\makebox[-0.8 mm][l]{/}{D}}_x\gamma _\rho \gamma _5
+\gamma _\rho \gamma _5\stackrel{\rightarrow}{\makebox[-0.8 mm][l]{/}{D}}_x)\psi (x)\psi (x_1)\bar{\psi}(x_2)) \nonumber \\
& &-{\lim _{x^{\prime }\rightarrow x}}(\partial _\lambda ^x-\partial _\lambda ^{x^{\prime }})T(\bar{\psi}(x^{\prime })\varepsilon ^{\lambda \mu \nu \alpha }
\gamma _5 U_P (x^{\prime },x)\psi (x)\psi (x_1)\bar{\psi}(x_2)) \nonumber \\
& &-{\lim _{x^{\prime }\rightarrow x}}(\partial _x^\alpha +\partial ^\alpha _{x^{\prime }})T(\bar{\psi}(x^{\prime })\sigma ^{\mu \nu }
U_P (x^{\prime },x)\psi (x)\psi (x_1)\bar{\psi}(x_2)),
\end{eqnarray}
where $\vec{D}_\mu =\vec{\partial}_\mu +igA_\mu $ and $\stackrel{\leftarrow}{D}_\mu=\stackrel{\leftarrow}{\partial } _\mu -igA_\mu $ are coveriant derivatives with $g=e$ and $A_\mu $ being the photon field in QED, and $g=g_c$ and $A_\mu =A_\mu ^aT^a$ in QCD with $A^a_\mu $ being the gluon field and $T^a$ being the generators of $SU(3)_c$ group. The wilson line $U_P (x^{\prime },x)=P\exp (-ig\int_x^{x^{\prime }}dy^\rho A_\rho (y))$ has been introduced in order that the current operators in the last term of Eqs.(1)-(3) be locally gauge invariant.

Taking into account the equations of motion for fermions with mass $ m$ : $(i{\stackrel{\rightarrow}{\makebox[-0.8 mm][l]{/}{D}} }-m)\psi =0$, $\bar{\psi}(i\stackrel{\leftarrow}{\makebox[-0.8 mm][l]{/}{D}}+m)=0$, which have the same form for both QED and QCD with the notation for covariant derivatives given above, we arrive at the identical relations among the transverse parts of the fermion's three-point functions in gauge theories(in coordinate space):
\begin{eqnarray}
& &\partial _x^\mu \left\langle 0\left| Tj^\nu (x)\psi (x_1)\bar{\psi}(x_2)\right| 0\right\rangle -\partial _x^\nu \left\langle 0\left| Tj^\mu (x)\psi (x_1)\bar{\psi}(x_2)\right| 0\right\rangle \nonumber \\
&=&i\sigma ^{\mu \nu }\left\langle 0\left| T\psi (x_1)\bar{\psi}(x_2)\right| 0\right\rangle \delta ^4(x_1-x)+i\left\langle 0\left| T\psi (x_1)\bar{\psi}(x_2)\right| 0\right\rangle \sigma ^{\mu \nu }\delta ^4(x_2-x)\nonumber \\
& &+2m\left\langle 0\left| T\bar{\psi}(x)\sigma ^{\mu \nu }\psi (x)\psi (x_1)\bar{\psi}(x_2)\right| 0\right\rangle \nonumber \\
& &+{\lim _{x^{\prime }\rightarrow x}}i(\partial _\lambda ^x-\partial _\lambda ^{x^{\prime }})\varepsilon ^{\lambda \mu \nu \rho }\left\langle 0\left| T\bar{\psi}(x^{\prime })\gamma _\rho \gamma _5U_P (x^{\prime },x)\psi (x)\psi (x_1)\bar{\psi}(x_2)\right| 0\right\rangle ,
\end{eqnarray}
\begin{eqnarray}
& &\partial _x^\mu \left\langle 0\left| Tj_5^\nu (x)\psi (x_1)\bar{\psi}(x_2)\right| 0\right\rangle -\partial _x^\nu \left\langle 0\left| Tj_5^\mu (x)\psi (x_1)\bar{\psi}(x_2)\right| 0\right\rangle 
\nonumber \\
&=&i\sigma ^{\mu \nu }\gamma _5\left\langle 0\left| T\psi (x_1)\bar{\psi}(x_2)\right| 0\right\rangle \delta ^4(x_1-x)-i\left\langle 0\left| T\psi (x_1)\bar{\psi}(x_2)\right| 0\right\rangle \sigma ^{\mu \nu }\gamma _5\delta ^4(x_2-x) \nonumber \\
& &+{\lim _{x^{\prime }\rightarrow x}}i(\partial _\lambda ^x-\partial _\lambda ^{x^{\prime }})\varepsilon ^{\lambda \mu \nu \rho }\left\langle 0\left| T\bar{\psi}(x^{\prime })\gamma _\rho U_P (x^{\prime },x)\psi (x)\psi (x_1)\bar{\psi}(x_2)\right| 0\right\rangle ,
\end{eqnarray}
and
\begin{eqnarray}
& &\partial _x^\mu \left\langle 0\left| Tj^{\nu \alpha}(x)\psi (x_1)\bar{\psi}(x_2)\right| 0\right\rangle -\partial _x^\nu \left\langle 0\left| Tj^{\mu \alpha }(x)\psi (x_1)\bar{\psi}(x_2)\right| 0\right\rangle 
\nonumber \\
&=&\varepsilon ^{\mu \nu \alpha \rho}\gamma_\rho \gamma _5\left\langle 0\left| T\psi (x_1)\bar{\psi}(x_2)\right| 0\right\rangle \delta ^4(x_1-x)-
\left\langle 0\left| T\psi (x_1)\bar{\psi}(x_2)\right| 0\right\rangle \varepsilon ^{\mu \nu \alpha \rho}\gamma _\rho \gamma _5\delta ^4(x_2-x) 	\nonumber \\
& &-{\lim _{x^{\prime }\rightarrow x}}(\partial _\lambda ^x-\partial _\lambda ^{x^{\prime }})\varepsilon ^{\lambda \mu \nu \alpha }\left\langle 0\left| T\bar{\psi}(x^{\prime })\gamma _5 U_P (x^{\prime },x)\psi (x)\psi (x_1)\bar{\psi}(x_2)\right| 0\right\rangle \nonumber \\
& &-{\lim _{x^{\prime }\rightarrow x}}(\partial ^\alpha _x+\partial ^\alpha _{x^{\prime }})\left\langle 0\left| T\bar{\psi}(x^{\prime })\sigma ^{\mu \nu }
U_P (x^{\prime },x)\psi (x)\psi (x_1)\bar{\psi}(x_2)\right| 0\right\rangle, 
\end{eqnarray}
where the vacuum expectation values have been used. Eqs.(4)-(6) are valid for both QED and QCD.

The transverse relations can be written in more clear and elegant form in the momentum space. By computing the Fourier transformation of Eqs.(4)-(6), respectively, and using the standard definition for the three-point functions in momentum space, we get the identical relations among the transverse parts of variant fermion's three-point functions in the momentum space:
\begin{eqnarray}
& &iq^\mu \Gamma _V^\nu (p_1,p_2)-iq^\nu \Gamma _V^\mu (p_1,p_2)\nonumber \\
&=&S_F^{-1}(p_1)\sigma ^{\mu \nu }+\sigma ^{\mu \nu }S_F^{-1}(p_2)+2m\Gamma _T^{\mu \nu }(p_1,p_2)\nonumber \\
& &+(p_{1\lambda }+p_{2\lambda })\varepsilon ^{\lambda \mu \nu \rho }\Gamma _{A\rho }(p_1,p_2),
\end{eqnarray}
\begin{eqnarray}
& &iq^\mu \Gamma _A^\nu (p_1,p_2)-iq^\nu \Gamma _A^\mu (p_1,p_2)\nonumber \\
&=&S_F^{-1}(p_1)\sigma ^{\mu \nu }\gamma _5-\sigma ^{\mu \nu }\gamma _5S_F^{-1}(p_2)+(p_{1\lambda }+p_{2\lambda })\varepsilon ^{\lambda \mu \nu \rho }\Gamma _{V\rho }(p_1,p_2),
\end{eqnarray}
and
\begin{eqnarray}
& &q^\mu \Gamma _T^{\nu \alpha}(p_1,p_2)+q^\nu \Gamma _T^{\alpha \mu } (p_1,p_2)+q^\alpha \Gamma _T^{\mu \nu }(p_1,p_2)\nonumber \\
&=&-S_F^{-1}(p_1)\varepsilon ^{\mu \nu \alpha \rho}\gamma _\rho \gamma _5+
\varepsilon ^{\mu \nu \alpha \rho}\gamma _\rho \gamma _5S_F^{-1}(p_2)+(p_{1\lambda }+p_{2\lambda })\varepsilon ^{\lambda \mu \nu \alpha }\Gamma _5(p_1,p_2),
\end{eqnarray}
where $q=p_1-p_2$, $\Gamma _V^\mu $, $\Gamma _A^\mu $, $\Gamma _T^{\mu \nu }$ and $\Gamma _5$ are the vector, the axial-vector, the tensor and the psudo-scalar vertex functions in momentum space, respectively, and $S_F(p_1)$ is the complete propagator of fermion. The third term of the left-hand side of Eq.(9) comes from the Fourier transformation of the last term in Eq.(6).
 
To understand the physics implication described by Eqs.(7)-(8) more clearly, we multiply both sides of Eqs.(7) and (8) by $iq_\nu $ and  then move the terms proportional to $q_\nu \Gamma _V^\nu $ and $q_\nu \Gamma _A^\nu $ into the right hand side of the equations, we thus have
\begin{eqnarray}
q^2\Gamma _V^\mu (p_1,p_2)&=&q^\mu (q_\nu \Gamma _V^\nu (p_1,p_2))+iS_F^{-1}(p_1)q_\nu \sigma ^{\mu \nu }+iq_\nu \sigma ^{\mu \nu }S_F^{-1}(p_2) \nonumber \\
 & &+2imq_\nu \Gamma _T^{\mu \nu }(p_1,p_2)+i(p_{1\lambda }+p_{2\lambda })q_\nu \varepsilon ^{\lambda \mu \nu \rho }\Gamma _{A\rho }(p_1,p_2),
\end{eqnarray}
\begin{eqnarray}
q^2\Gamma _A^\mu (p_1,p_2)&=&q^\mu (q_\nu \Gamma _A^\nu (p_1,p_2))+iS_F^{-1}(p_1)q_\nu \sigma ^{\mu \nu }\gamma _5-iq_\nu \sigma ^{\mu \nu }\gamma _5S_F^{-1}(p_2)\nonumber \\
& &+i(p_{1\lambda }+p_{2\lambda })q_\nu \varepsilon ^{\lambda \mu \nu \rho }\Gamma _{V\rho }(p_1,p_2).
\end{eqnarray}
Now writing the full vertices, $\Gamma _V^\mu $ and $\Gamma _A^\mu $, as
\begin{equation}
\Gamma _V^\mu (p_1,p_2)=\Gamma _{V(L)}^\mu (p_1,p_2)+\Gamma _{V(T)}^\mu (p_1,p_2),
\end{equation}
\begin{equation}
\Gamma _A^\mu (p_1,p_2)=\Gamma _{A(L)}^\mu (p_1,p_2)+\Gamma _{A(T)}^\mu (p_1,p_2),
\end{equation}
we then obtain from Eqs.(9) and (10):
\begin{equation}
\Gamma _{V(L)}^\mu (p_1,p_2)=q^{-2}q^\mu (q_\nu \Gamma _V^\nu (p_1,p_2)),
\end{equation}
\begin{eqnarray}
\Gamma _{V(T)}^\mu (p_1,p_2)&=&q^{-2}q_\nu [iS_F^{-1}(p_1)\sigma ^{\mu \nu }+i\sigma ^{\mu \nu }S_F^{-1}(p_2)+2im\Gamma ^{\mu\nu}_T (p_1, p_2) \nonumber \\
& &+i(p_{1\lambda }+p_{2\lambda })\varepsilon ^{\lambda \mu \nu \rho }\Gamma _{A\rho }(p_1,p_2)],
\end{eqnarray}
and
\begin{equation}
\Gamma _{A(L)}^\mu (p_1,p_2)=q^{-2}q^\mu (q_\nu \Gamma _A^\nu (p_1,p_2)),
\end{equation}
\begin{equation}
\Gamma _{A(T)}^\mu (p_1,p_2)=q^{-2}q_\nu [iS_F^{-1}(p_1)\sigma ^{\mu \nu }\gamma _5-i\sigma ^{\mu \nu }\gamma _5S_F^{-1}(p_2)+i(p_{1\lambda }+p_{2\lambda })\varepsilon ^{\lambda \mu \nu \rho }\Gamma _{V\rho }(p_1,p_2)].
\end{equation}
By using the antisymmetry property of $\sigma ^{\mu \nu }$ and $\varepsilon ^{\lambda \mu \nu \rho }$, it is easy to check that $q_\mu \Gamma _{V(T)}^\mu =0$ and $q_\mu \Gamma _{A(T)}^\mu =0$, that is,  $\Gamma _{V(T)}^\mu $ and $\Gamma _{A(T)}^\mu $ are indeed the transverse components. Note that $\Gamma _{V(T)}^\mu $ and $\Gamma _{A(T)}^\mu $ are just the right side of Eq.(7) and Eq.(8), respectively, except the factor $iq^{-2}q_\nu $. Therefore Eqs.(7) and (8)(and the corresponding expressions in coordinate space, Eqs.(4)
and (5)) describe respectively the relations among the transverse part of the vector and the axial-vector vertex functions and other Green's functions. Eqs.(4) and (7) show that the transverse part of the vector vertex function is related to the inverse of the fermion propagator, the tensor vertex function and the axial-vector vertex function, while Eqs.(5) and (8) show that the transverse part of the axial-vector vertex function is related to the inverse of the fermion propagator and the vector vertex function. Thus, the transverse parts of variant vertex functions are coupled each other. As a result, the full vector and the full axial-vector vertex functions are also coupled each other and form a set of coupled equations, which is described by Eqs.(10) and (11). The constraint relation (9) can be similarly discussed.

In Eqs.(10)-(16), $q_\mu \Gamma _V^\mu $ and $q_\mu\Gamma _A^\mu $ satisfy the normal Ward-Takahashi identities:
\begin{equation}
q_\mu \Gamma _V^\mu (p_1,p_2)=S_F^{-1}(p_1)-S_F^{-1}(p_2),
\end{equation}
\begin{equation}
q_\mu \Gamma _A^\mu (p_1,p_2)=S_F^{-1}(p_1)\gamma _5+\gamma _5S_F^{-1}(p_2)-2im\Gamma _5(p_1,p_2),
\end{equation}
 Besides these two identities, we also need the W-T identity for the tensor vertex function, $q_\nu \Gamma _T^{\mu \nu }$. By the procedure similar to that of deriving Eqs.(18) and (19), we find
\begin{equation}
iq_\nu \Gamma _T^{\mu \nu }(p_1,p_2)=S_F^{-1}(p_1)\gamma ^\mu +\gamma ^\mu S_F^{-1}(p_2)+2m\Gamma _V^\mu (p_1,p_2)+(p_1^\mu +p_2^\mu )\Gamma _S(p_1,p_2),
\end{equation}
where $\Gamma _S$ is the scalar vertex function. 

Now we have the normal Ward-Takahashi identities describing the longitudinal part of the three-point vertex functions, given by Eqs.(18)-(20), and the identical relations among transverse parts of variant three-point vertex functions given by Eqs.(7)-(9), which form a complete set of constraint relations(Ward-Takahashi type's relations) for variant three-point vertex functions in gauge theories. As a consequence, we can consistently write the full vertex functions, $\Gamma _V^\mu $, $\Gamma _A^\mu $ and $\Gamma _T^{\mu \nu}$ in terms of this complete set of relations without any Ans\"{a}tze. 

In fact, by substituting Eq.(11) into Eq.(10) and using Eqs.(18)-(20), it is not difficult to obtain the full vector-vertex function:
\begin{equation}
\Gamma _V^\mu (p_1,p_2)=\Gamma _{V(L)}^\mu (p_1,p_2)+\Gamma _{V(T)}^\mu (p_1,p_2)
\end{equation}
with
\begin{equation}
\Gamma _{V(L)}^\mu (p_1,p_2)=q^{-2}q^\mu (S_F^{-1}(p_1)-S_F^{-1}(p_2)),
\end{equation}
\begin{eqnarray}
\Gamma _{V(T)}^\mu (p_1,p_2)&=&(q^2+(p_1+p_2)^2-4m^2-((p_1+p_2)\cdot q)^2q^{-2})^{-1}\nonumber \\
& &\times \{(S_F^{-1}(p_1)-S_F^{-1}(p_2))[4m^2q^\mu +q^\mu ((p_1+p_2)\cdot q)^2q^{-2}-(p_1^\mu +p_2^\mu )(p_1+p_2)\cdot q]q^{-2}\nonumber \\
& &+(S_F^{-1}(p_1)+S_F^{-1}(p_2))(p_1^\mu +p_2^\mu -q^\mu (p_1+p_2)\cdot qq^{-2})\nonumber \\
& &+iS_F^{-1}(p_1)\sigma ^{\mu \nu }q_\nu +i\sigma ^{\mu \nu }q_\nu S_F^{-1}(p_2)\nonumber \\
& &+i(S_F^{-1}(p_1)\sigma ^{\mu \lambda }-\sigma ^{\mu \lambda }S_F^{-1}(p_2))(p_{1\lambda }+p_{2\lambda })\nonumber \\
& &+i(S_F^{-1}(p_1)\sigma ^{\lambda \nu }-\sigma ^{\lambda \nu }S_F^{-1}(p_2))q_\nu (p_{1\lambda }+p_{2\lambda })q^\mu q^{-2}\nonumber \\
& &-i(S_F^{-1}(p_1)\sigma ^{\mu \nu }-\sigma ^{\mu \nu }S_F^{-1}(p_2))q_\nu (p_1+p_2)\cdot qq^{-2}\nonumber \\
& &+2m(S_F^{-1}(p_1)\gamma ^\mu +\gamma ^\mu S_F^{-1}(p_2)-q^\mu (p_1+p_2)\cdot qq^{-2}\Gamma _S(p_{1,}p_2))\},
\end{eqnarray}
Similarly, by substituting Eqs.(21)-(23) into Eq.(11) and using Eq.(19), we can write the full axial-vector vertex function as
\begin{equation}
\Gamma _A^\mu (p_1,p_2)=\Gamma _{A(L)}^\mu (p_1,p_2)+\Gamma _{A(T)}^\mu (p_1,p_2)
\end{equation}
with
\begin{equation}
\Gamma _{A(L)}^\mu (p_1,p_2)=q^{-2}q^\mu (S_F^{-1}(p_1)\gamma _5+\gamma _5S_F^{-1}(p_2)-2im\Gamma _5(p_{1,}p_2)),
\end{equation}
\begin{eqnarray}
\Gamma _{A(T)}^\mu (p_1,p_2)&=&(q^2+(p_1+p_2)^2-4m^2-((p_1+p_2)\cdot q)^2q^{-2})^{-1}\nonumber \\
& &\times \{i(S_F^{-1}(p_1)\gamma _5\sigma ^{\mu \nu }-\sigma ^{\mu \nu }\gamma _5S_F^{-1}(p_2))q_\nu (1-4m^2q^{-2})\nonumber \\
& &-i(S_F^{-1}(p_1)\gamma _5\sigma ^{\mu \nu }+\sigma ^{\mu \nu }\gamma _5S_F^{-1}(p_2))q_\nu (p_1+p_2)\cdot qq^{-2}\nonumber \\
& &+i(S_F^{-1}(p_1)\gamma _5\sigma ^{\mu \lambda }+\sigma ^{\mu \lambda }\gamma _5S_F^{-1}(p_2))(p_{1\lambda }+p_{2\lambda })\nonumber \\
& &+i(S_F^{-1}(p_1)\gamma _5\sigma ^{\lambda \nu }+\sigma ^{\lambda \nu }\gamma _5S_F^{-1}(p_2))q_\nu (p_{1\lambda }+p_{2\lambda })q^\mu q^{-2}\nonumber \\
& &+i(S_F^{-1}(p_1)\gamma _5\sigma ^{\lambda \nu }-\sigma ^{\lambda \nu }\gamma _5S_F^{-1}(p_2))q_\nu (p_{1\lambda }+p_{2\lambda }) (p_1^\mu +p_2^\mu -q^\mu (p_1+p_2)\cdot qq^{-2})q^{-2}\nonumber \\
& &+2im(S_F^{-1}(p_1)\gamma _\rho +\gamma _\rho S_F^{-1}(p_2))\varepsilon ^{\lambda \mu \nu \rho }(p_{1\lambda }+p_{2\lambda })q_\nu \}.
\end{eqnarray}
We see that the full vector and the full axial-vector vertex functions are now expressed in terms of two-point functions (fermion propagators) and the scalar and the pseudo-scalar vertex functions, respectively. In the chiral limit, the latter two vertex functions will have no contribution. 

Finally, by using Eqs.(9) and (20) we can get the full tensor vertex
function:
\begin{eqnarray}
q^2\Gamma _T^{\mu \nu}(p_1,p_2)&=&
iS_F^{-1}(p_1)(q^\mu \gamma ^\nu -q^\nu \gamma ^\mu)+
i(q^\mu \gamma ^\nu -q^\nu \gamma ^\mu)S_F^{-1}(p_2)\nonumber \\
& &+2im(q^\mu \Gamma _V^\nu (p_1,p_2)-q^\nu \Gamma _V^\mu (p_1,p_2))+
i[q^\mu (p_1^\nu +p_2^\nu ) -q^\nu (p_1^\mu+p_2^\mu)]\Gamma _S(p_1,p_2)
\nonumber \\
& &-S_F^{-1}(p_1)\varepsilon ^{\mu \nu \alpha \rho }q_\alpha \gamma _\rho \gamma _5 + \varepsilon ^{\mu \nu \alpha \rho }q_\alpha \gamma _\rho \gamma _5S_F^{-1}(p_2) + (p_{1\lambda }+p_{2\lambda })q_\alpha \varepsilon ^{\lambda \alpha \mu \nu }\Gamma _5(p_1,p_2),
\end{eqnarray}
where $\Gamma _V^{\mu (\nu)}$ is given by Eqs.(21)-(23).

Some comments are given as follows:

(i) The constraint relation for the transverse part of the vector vertex function given by Eqs.(4) and (7) involves the mass term arising from the equations of motion. This situation is similar to the case for the normal Ward-Takahashi identity for the axial-vector vertex(see Eq.(19)) due to the partial conservation of axial-vector current(PCAC). On the contrary, the constraint relations for the transverse part of the axial-vector vertex function given by Eqs.(5) and (8) have no mass term, which is similar to
the case for the W-T identity for the vector vertex function(see Eq.(18)).

(ii) The transverse relations for the three-point functions given by Eqs.(4)-(9) have been derived in the classical QED and QCD. It remains to show if these identical relations are modified by higher-order correction terms in perturbation theory. It is well-known that the normal W-T identity for the axial-vector vertex, given by Eq.(19), 
is modified due to the Adler-Bell-Jackiw anomaly[8]. As a result, the W-T identity for axial-vector vertex function, given by Eq.(19), should add the anomaly term contribution. By applying the approach for deriving ABJ anomaly [8][9] to the present case, we find that the ABJ anomaly does not contribute to the transverse relations, Eqs.(4)-(9). The modification by higher-order correction does happen to the tensor vertex term due to the renormalization of the tensor current operator, which leads to the appearance of anomalous dimension in the tensor vertex term. 

(iii) In the chiral limit the full vertex functions, $\Gamma _V^\mu $ and $\Gamma _A^\mu $, are expressed in terms of the two-point functions (fermion propagators) only. Applying these results to the Dyson-Schwinger equations will lead to that these equations form a closed system for the two-point functions in QED and classical QCD case. In QCD case we usually consider the vertex functions involving the current operators,
$j^{\mu}(x)=\bar{\psi}(x) \gamma^{\mu} \frac{\lambda ^a}{2} \psi(x)$
and
$j^{\mu}_5(x)=\bar{\psi}(x) \gamma^{\mu} \gamma_5\frac{\lambda ^a}{2} \psi(x)$,
respectively, where $\lambda ^a$ are flavor generators. In such case,
the relative results of present work will be simply modified just by putting
$\frac{\lambda ^a}{2}$ into the suitable position in each term of the
correspoding identical relations.
For the case of effective QCD with Faddeev-Popov ghost fields, there appear to be more vertices. The constraint relations among transverse parts of these new vertices and the full vertices need to be studied further.

In summary, we have derived the Ward-Takahashi type's identical relations among the transverse parts of variant fermion's three-point vertex functions in the
coordinate space as well as in the momentum space. These transverse relations together with the normal (longitudinal) Ward-Takahashi identities form a complete set of W-T type's constraint relations for three-point vertex functions. As a consequence, the full vector, the full axial-vector and the full tensor vertex functions in the momentum space have been consistently and exactly deduced in terms of this complete set of W-T relations. It can be expected that these full vertex functions are very useful to the nonperturbative study of gauge field theroy using the Dyson-Schwinger equation approach and its application to hadronic physics.

\section*{Acknowledgments}

The author is very grateful to Y.Takahashi and F.C.Khanna for useful discussions during his visit at the University of Alberta, where the part of this work was done. This work is supported in part by the National Natural Science Foundation of China.

\end{document}